\documentclass[reprint,aps,ams,amsmath,prx,twocolumn,superscriptaddress,longbibliography]{revtex4-1}
\usepackage{graphics}
\usepackage{graphicx}
\usepackage{epsfig}
\usepackage{amsmath}
\usepackage{amssymb}
\usepackage{amsfonts}
\usepackage{bm}
\usepackage{color}
\usepackage{bbm}
\usepackage[bookmarks=false,linkcolor=blue,urlcolor=blue,colorlinks,citecolor=blue]{hyperref}
\usepackage[normalem]{ulem}
\usepackage{comment}
\usepackage{soul}

\begin{document}

\title{Thermal transfer enhancement by hydrodynamic plasmons in electron bilayers}

\author{Dmitry Zverevich}
\affiliation{Department of Physics, University of Wisconsin-Madison, Madison, Wisconsin 53706, USA}

\author{A. V. Andreev}
\affiliation{Department of Physics, University of Washington, Seattle, Washington 98195, USA}

\author{Alex Levchenko}
\affiliation{Department of Physics, University of Wisconsin-Madison, Madison, Wisconsin 53706, USA}

\date{August 14, 2023}

\begin{abstract}
We develop a theory of heat transfer induced by thermal charge fluctuations in two-dimensional electron double layers. 
We consider pristine systems comprised of identical layers, and focus on the regime of sufficiently high temperatures and interlayer distances $d$, where the relevant charge fluctuations may be described using the hydrodynamic approach. In this limit heat transfer is dominated by the plasmon resonances. For systems with Galilean-invariant electron dispersion the interlayer thermal conductance $\varkappa$ is proportional to the kinematic viscosity of the electron liquid, and decreases as $1/d^4$. In the absence of Galilean invariance $\varkappa \propto \sigma/d^3$, where $\sigma$ is the intrinsic conductivity of the liquid. This strong enhancement can be traced to a drastically different broadening of plasmon resonances in systems with and without Galilean invariance.          
\end{abstract}

\maketitle

\section{Introduction}

The Stefan-Boltzmann law \cite{Stefan,Boltzmann} of radiative heat flux emitted by a blackbody per unit area, $W=\sigma_{\text{SB}} T^4$, was explained by Planck in the first quantum theory \cite{Planck} to be determined only by its temperature $T$ with the universal constant $\sigma_{\text{SB}}=\pi^2/60c^2$, where $c$ is the speed of light \footnote{In this paper we work in the units where Planck and Boltzmann constants are set to unity $\hbar=k_B=1$}. This law holds asymptotically at large distances from the body. In a geometry where two bodies are separated by a finite distance heat between them can also be transferred by nonradiative evanescent modes of the electromagnetic field \cite{Hargreaves,PolderVanHove,Levin,Pendry,Loomis}. At sufficiently small separations $d$ the contribution of this near-field energy transfer (NFET) may greatly exceed that of blackbody. Physically, NFET arises because electromagnetic fields produced by fluctuating charges and currents in one system induce currents in the other. In the presence of a temperature difference the energy dissipated by these fluctuating fields in the other part of the structure leads to a net heat transfer.
 
In metallic systems charge fluctuations may propagate in the form of collective plasmon excitations. Mahan \cite{Mahan} pointed out that surface plasmons in metal bilayers could lead to enhanced interlayer heat transfer. While in bulk samples the plasmons are gapped, in systems of reduced dimensionality they have a  gapless spectrum, leading to further enhancement of NFET. Stimulated by the advances in the field of plasmonics \cite{Maccaferri} and advent of two-dimensional (2D) van der Waals materials \cite{Geim}, plasmonic tuning of NFET attracted considerable attention recently \cite{Svetovoy,Ilic,Rodriguez,Raikh,Yu,Jiang,Zhao,Kamenev,Zhang,Wang,Ying,Hekking,Basko}. There is a large number of recent experiments \cite{Kim,Song,St-Gelais,Yang} which report observations of excess of heat flux in circuits between closely spaced electronic nanostructures. For recent reviews, see Refs. \cite{Volokitin,Bimonte,BenAbdallah} and references therein. 

The broadening of the plasmon pole plays a crucial effect on the magnitude of plasmon enhancement of NFET. The existing calculations  focused on the collisionless regime, where the plasmon attenuation arises from Landau damping and may be evaluated using the random-phase approximation (RPA). Here we focus on the opposite, collision-dominated regime, in which the rate of electron-electron collisions, $\gamma_{ee}$, exceeds the plasmon damping rate. Furthermore, we specialize to the hydrodynamic regime \cite{Narozhny,Scaffidi}, in which $\gamma_{ee}$ exceeds not only the damping rate, but the frequencies of charge fluctuations responsible for energy transfer. Since the characteristic wave vectors of such fluctuations may be estimated as $1/d$ the  conditions applicability of the hydrodynamic description can be satisfied in modern high mobility devices at intermediate temperatures. 

Importantly, we find that the presence or absence of Galilean invariance of the electron liquid drastically affects the plasmon damping rate and consequently the magnitude of NFET. Specifically, in Galilean invariant systems the plasmon damping rate depends on the wavevector $q$ as $ \nu q^2$, where $\nu$ is the kinematic viscosity. In the absence of Galilean invariance the electron liquid acquires a nonvanishing intrinsic conductivity $\sigma$ and the plasmons decay with the Maxwell relaxation rate $2\pi \sigma q $. The different momentum dependence of the plasmon attenuation rate produces a different dependence of the NFET interlayer  thermal conductance $\varkappa$ on $d$. In the Galilean invariant case $\varkappa \propto \nu/d^4$, while in the absence of Galilean invariance we get (modulo $\ln d$ terms)  $\varkappa \propto \sigma/d^3$. It is interesting to note that in the absence of Galilean invariance the $1/d^3$ dependence  of $\varkappa$ persists not only at a nonzero electron density, but also at double charge neutrality, where the plasmon resonances do not exist. 

The presentation is organized as follows. In Sec. \ref{sec:hydro} we summarize the theory of hydrodynamic fluctuations in electron liquids that form the basis of our analysis. Conceptually it parallels with the methods of fluctuation electrodynamics developed by Rytov \cite{Rytov} as we use macroscopic equations of motion for electron fluid with inclusion of random Langevin fluxes. The latter are supplemented by the fluctuation-dissipation relation via kinetic dissipative coefficients of the pristine fluid such as viscosity and thermoelectric matrix. In Sec. \ref{sec:nfht} we use the Ehrenfest theorem to derive an expression for the NFET flux. It provides an elegant methodological step that enables us to make a connection to the commonly used Caroli formula \cite{Yu,Jiang,Zhao,Kamenev,Zhang,Wang,Ying}. Next we analyze our general expression in several cases of interest. This includes the examples of Galilean invariant electron systems, such as electrons in parabolic partially occupied quantum wells of semiconductor heterostructures and generic liquids without Galilean invariance, such as Dirac fluid in graphene. Furthermore, the tunability of graphene-based systems motivates exploration of the NFET in the crossover from the high doping to the dual charge neutrality. We close in Sec. \ref{sec:summ} with the summary of main results, comparison to related works, and also provide relevant estimates.     

\begin{figure}[t!]
\centering
\includegraphics[width=3in]{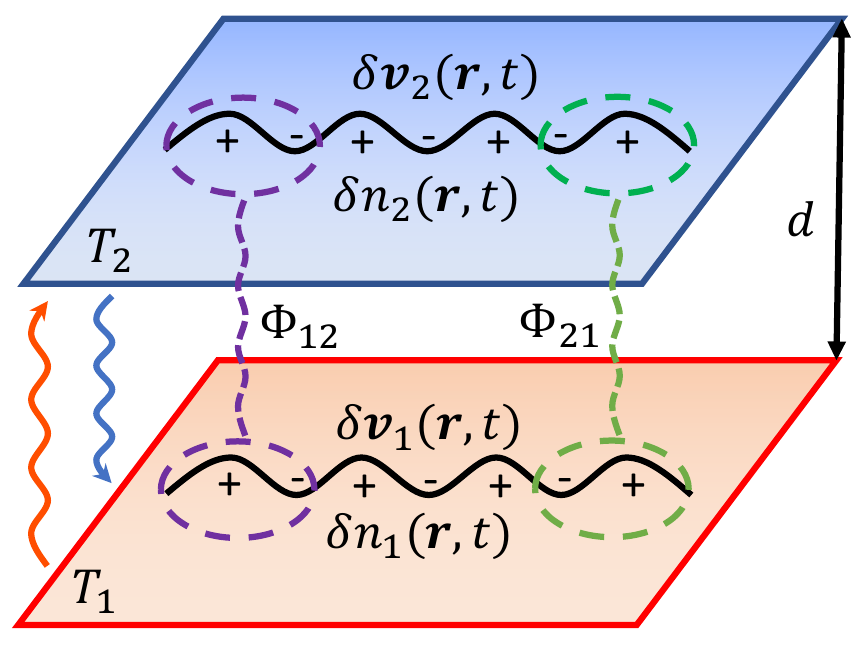}
 \caption{Schematic representation of an electron double layer under consideration. Thermally-driven electron density fluctuations in each layer $\delta n_{1,2}(\bm{r},t)$  depicted by propagating plasmon waves  are coupled by the interlayer Coulomb potential $\Phi_{12}$. By virtue of the continuity equation the conjugated hydrodynamic velocity fluctuations are labeled by $\delta\bm{v}_{1,2}(\bm{r},t)$. Each layer is kept at different temperatures $T_{1,2}$ and wavy lines pointing up or down represent the heat fluxes between the layers.}
\label{Fig-EDL}
\end{figure}

\section{Hydrodynamic fluctuations}\label{sec:hydro}

We consider a planar geometry of two conducting two-dimensional layers kept at different temperatures as shown in Fig. \ref{Fig-EDL}. For simplicity we consider identical layers. We assume that at the relevant frequencies and wavevectors the electron liquids in each layer can be described using the hydrodynamic approach. This requires that intralayer mean free path due to electron-electron collisions $l(T)$ is shorter than the interlayer separation, $l\ll d$. The latter sets a typical momentum transfer $q\sim1/d$ from one layer to the other thus leading to the momentum relaxation. The electrons in the two layers interact via the dynamically screened Coulomb potential. As a result, density fluctuations become strongly coupled. This produces two branches of collective modes in the bilayer, the optical and acoustic plasmons, corresponding to the  electron density oscillations, which are symmetric and antisymmetric in the layer index. We show below that thermal fluctuations of these plasmon resonances enhance NFET. To capture this physics we use the theory of hydrodynamic fluctuations formulated by Landau and Lifshitz \cite{LL} with an extension to include the Coulomb law in charged fluids. 

Quite generally, the hydrodynamic equations have the form of continuity equations, expressing  conservation of particle number (charge), energy, and momentum \cite{LL-V6}.  It further assumes local thermal equilibrium, so that the state of the liquid is characterized by the local equilibrium parameters, such as the hydrodynamic velocity $\bm{v}$, temperature $T$, chemical potential $\mu$, and pressure $P(n,T)$ that defines an equation of state. Noting that the entropy production is quadratic in deviations from equilibrium, it may be neglected for the purpose of studying linear transport. Therefore, in this approximation, energy conservation law can be replaced by the continuity equation for the entropy current. It is convenient to combine the continuity equations for the particle and entropy currents into a single equation 
\begin{equation}\label{eq:dt-x}
\partial_t\vec{x}=-\bm{\nabla}\cdot\vec{\bm{J}}.
\end{equation}
Here, we introduced the column vector $\vec{x}(\bm{r},t)$ that consists of the number density $n$ and the entropy density $s$. Similarly, we introduced the corresponding column vector $\vec{\bm{J}}(\bm{r},t)$ of the particle density current $\bm{j}_n$ and the entropy density current $\bm{j}_s$,  where the boldface letters denote the usual spatial vectors as opposed to column vectors 
\begin{equation}\label{eq:x-J-X}
\vec{x}=\left(\begin{array}{c}n \\ s\end{array}\right),\quad \vec{\bm{J}}=\left(\begin{array}{c}\bm{j}_n \\ \bm{j}_s\end{array}\right), 
\quad \vec{\bm{X}}=\left(\begin{array}{c}-e\bm{\mathcal{E}} \\ \bm{\nabla}T\end{array}\right).
\end{equation}
For future use we also introduced a column vector of thermodynamic forces $\vec{\bm{X}}$, which consists of the electromotive force (EMF) $e\bm{\mathcal{E}}$ and the local temperature gradient, which are thermodynamically conjugate~\cite{LL-V5} to the corresponding densities $\vec{x}$. 

Writing the momentum density in the form $\bm{p} = \rho \bm{v}$, where $\rho$ is the mass density~\footnote{In Galilean invariant systems $\rho =m n$, where $m$ is the band mass. In the absence of Galilean invariance $\rho$ is defined as the proportionality coefficient between the momentum density and the hydrodynamic velocity. For example in Dirac liquids with linear energy spectrum $\varepsilon = v p$, we have $\rho \sim T^3/v^4$.}, we write the momentum evolution equation in the form of Newton's second law,
\begin{equation}\label{eq:dt-p}
\rho \partial_t\bm{v}=-\bm{\nabla}\cdot\hat{\Pi}-en\bm{\nabla}\Phi.
\end{equation}
Here the electric potential $\Phi$ is related to the electron density $n$ by the Poisson equation. It defines the EMF $e\bm{\mathcal{E}}=-\bm{\nabla}(\mu+e\Phi)$ in Eq. \eqref{eq:x-J-X}. The form of Eq. \eqref{eq:dt-p} reflects the fact that the flow of momentum comprises both long-range Coulomb interactions between electrons and also local fluxes. The corresponding flux tensor $\hat{\Pi}\equiv\Pi_{ij}$ has the form \cite{LL-V6}
\begin{equation}
\Pi_{ij}=P\delta_{ij}-\Sigma_{ij}.
\end{equation}
It includes local hydrodynamic pressure $P$ and viscous stress tensor $\Sigma_{ij}$.
In the hydrodynamic approximation the fluxes of conserved quantities are expanded to first order in the gradients of equilibrium parameters.
This leads to the standard expression 
\begin{equation}\label{eq:stress}
\Sigma_{ij}=\eta(\partial_iv_j+\partial_jv_i)+(\zeta-\eta)\delta_{ij}\partial_kv_k+\Xi_{ij},
\end{equation}
where $\eta$ and $\zeta$ are, respectively, the shear and bulk viscosities. In this framework hydrodynamic fluctuations rendered by random forces are described by including Langevin sources $\Xi_{ik}(\bm{r},t)$, whose correlation function is given by \cite{LL}
\begin{align}\label{eq:zeta-zeta}
&\left\langle\Xi_{ik}(\bm{r},t)\Xi_{lm}(\bm{r}',t')\right\rangle= \nonumber \\ 
&2T  \delta(\bm{r}-\bm{r}')\delta(t-t')  
[\eta(\delta_{il}\delta_{km}+\delta_{im}\delta_{kl})+(\zeta-\eta)\delta_{ik}\delta_{lm}].
\end{align}

To close the system of hydrodynamic equations we also need expressions for currents $\vec{\bm{J}}$. In Galilean-invariant liquids, densities of both particle and entropy currents are uniquely determined by the local hydrodynamic velocity $\bm{v}(\bm{r},t)$. More generally, in the absence of Galilean invariance, additional dissipative contributions to currents arise \cite{Aleiner,Lucas,Li}. 
We thus have 
\begin{equation}\label{eq:J}
\vec{\bm{J}}=\vec{x}\bm{v}-\hat{\Upsilon}\vec{\bm{X}}+\vec{\bm{I}}.
\end{equation}
The first term on the right-hand side of the above equation is the usual convective part of the current that describes transport of charge and entropy in the macroscopic flow of the fluid with the hydrodynamic velocity $\bm{v}$. The second term captures dissipative transport of charge and heat relative to the fluid in response to driving forces $\vec{\bm{X}}$.  
The matrix of kinetic coefficients $\hat{\Upsilon}$ characterizes the dissipative properties of the electron liquid. It is defined by
\begin{equation}\label{eq:Upsilon}
\hat{\Upsilon}=\left(\begin{array}{cc}\sigma/e^2 & \gamma/T \\ \gamma/T & \kappa/T\end{array}\right).
\end{equation}
The diagonal elements of this matrix are given by the intrinsic conductivity $\sigma$ and the thermal conductivity $\kappa$. The off-diagonal element is the thermoelectric coefficient $\gamma$. The matrix form of $\hat{\Upsilon}$ respects the Onsager symmetry condition. In the Galilean invariant system $(\sigma,\gamma)\to0$. We should also note that in systems with parabolic and linear spectra bulk viscosity is known to vanish, $\zeta\to0$ \cite{LL-V10}. The fluctuation-dissipation relation dictates that the current must contain fluctuating components in conjunction with the corresponding dissipative transport coefficients. This is reflected by the third term on the right-hand side of Eq.~\eqref{eq:J}. It captures the thermally driven spatial and temporal fluctuations of Langevin currents $\vec{\bm{I}}(\bm{r},t)$. Their variances are defined by the dissipative matrix $\hat{\Upsilon}$ in a usual way \cite{LL-V9,Kogan},
\begin{equation}\label{eq:J-J}
\left\langle\vec{\bm{I}}(\bm{r},t)\otimes\vec{\bm{I}}^{\mathbb{T}}(\bm{r}',t')\right\rangle=2T\hat{\Upsilon}\delta(\bm{r}-\bm{r}')\delta(t-t'). 
\end{equation}
The notation $\vec{A}\otimes\vec{B}^{\mathbb{T}}$ is used to denote the direct product of two vectors and symbol $\mathbb{T}$ means vector transposition. 

In the geometry of the bilayer this system of equations needs to be replicated for each layer. In other words, we have to assign fluctuating densities $\delta n_{1,2}$ and velocities $\delta\bm{v}_{1,2}$ for each layer that are coupled together in the momentum balance equation through the Coulomb potential. This hydrodynamic approach was recently used to address the related problems of the Coulomb and thermal drag \cite{Apostolov,Patel,Pesin,Levchenko} and can be readily applied to the NFET effect as we show in the next section.   

\section{Near-field energy transfer}\label{sec:nfht}

The near-field thermal conductance can be straightforwardly evaluated with the help of the formalism presented above. For that purpose, we calculate the energy flux from layer $1$ to layer $2$ by computing the work per unit time done by the density fluctuations in  layer $1$ on the electrons in layer $2$. Using Ehrenfest's theorem \cite{LL-V5}, we can write the heat flux per unit area in the form \footnote{Note that we we do not account for electron tunneling between the layers. For example, in graphene devices encapsulated with hexagonal boron nitride, this is fairly safe assumption.}
\begin{equation}
W=\frac{e}{2}\left\langle\delta n_2\partial_t\delta\Phi_2-\delta n_1\partial_t\delta\Phi_1\right\rangle,
\end{equation}
where $\langle \ldots \rangle $ denotes thermal average.
To evaluate this average it is useful to pass to the Fourier space for all fluctuating quantities, $(\delta n,\delta\bm{v},\delta\Phi)\propto \exp(i\bm{qr}-i\omega t)$. This enables us  to express the heat flux  
\begin{equation}\label{eq:W}
W=\frac{1}{2}\int\frac{\omega d\omega d^2q}{(2\pi)^3}\left(\frac{2\pi e^2}{\epsilon q}\right)e^{-qd}\, \Im D(\bm{q},\omega), 
\end{equation}
in terms of the imaginary part of the dynamical structure factor of density fluctuations
\begin{equation}\label{eq:D}
D(\bm{q},\omega)=\left\langle\delta n_+(-\bm{q},-\omega)\delta n_-(\bm{q},\omega)\right\rangle.
\end{equation}
To arrive at Eq.~\eqref{eq:W} we took the Coulomb potential in the form  
\begin{equation}\label{eq:delta-Phi}
\delta\Phi_{1}(\bm{q},\omega)=\left(\frac{2\pi e}{\epsilon q}\right)\left[\delta n_1(\bm{q},\omega)+\delta n_2(\bm{q},\omega)e^{-qd}\right],
\end{equation}
where $\epsilon$ is the dielectric constant of the material surrounding the electron layers, and made an additional linear transformation to the symmetrized basis of density fluctuations $\delta n_\pm=\delta n_1\pm\delta n_2$. The knowledge of $W$ naturally leads to the NFET thermal conductance, which we will determine to linear order in the temperature difference between the layers $\Delta T=T_1-T_2$. We thus introduce 
\begin{equation}\label{eq:NFHT}
\varkappa(T)=\lim_{\Delta T\to 0}\frac{W}{\Delta T}. 
\end{equation}
The technical task now is to derive an analytical formula for the density correlation function. This can be done by solving linearized hydrodynamic equations and performing thermal averages over the Langevin sources.    

\begin{figure}[t!]
  \centering
  \includegraphics[width=3.25in]{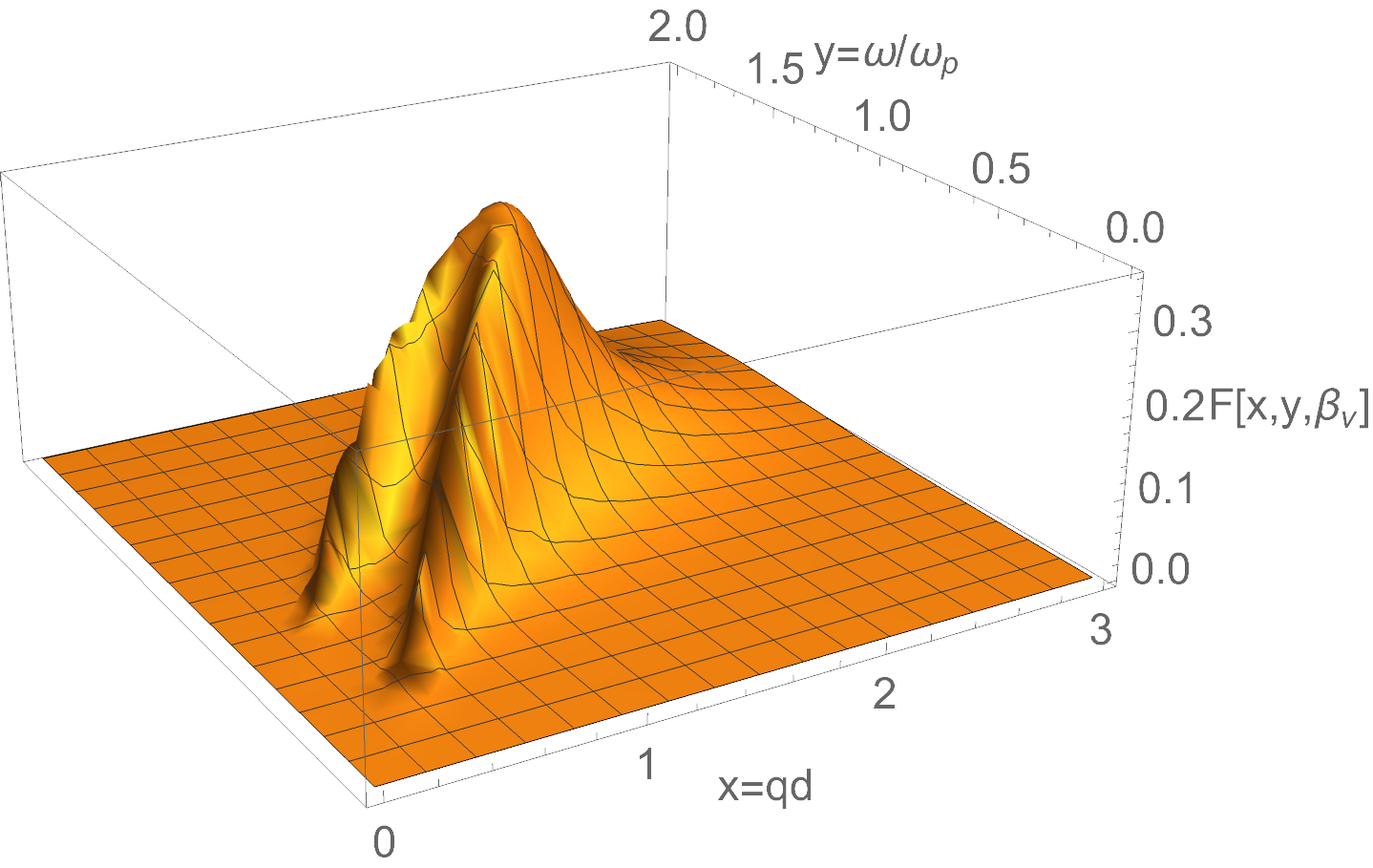}
  \includegraphics[width=3.25in]{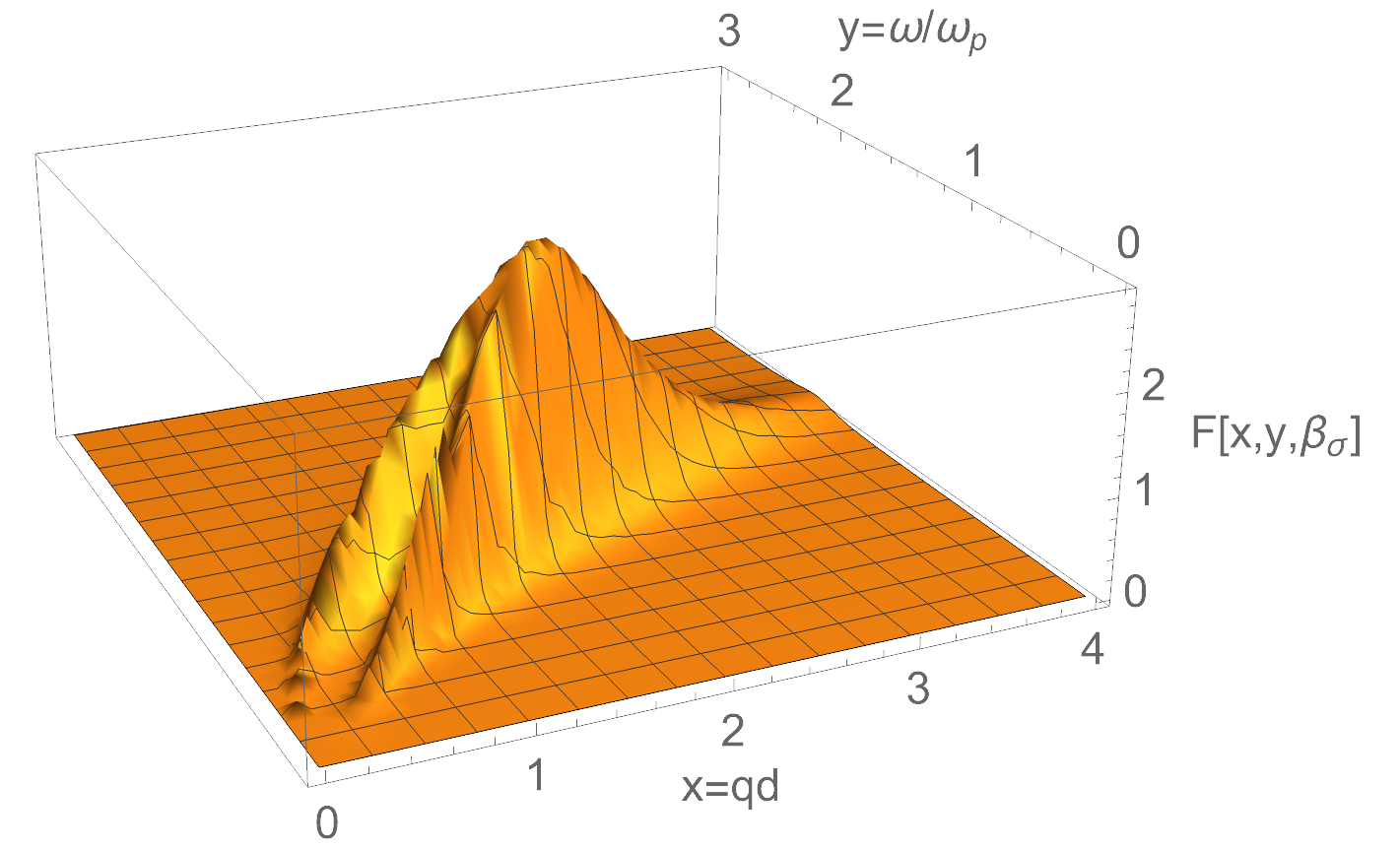}
  \caption{Plot shows the integrand of Eq. \eqref{eq:W} denoted as $F(x,y,\beta)$ in the dimensionless units of momenta $x=qd$ and frequency $y=\omega/\omega_p$. 
  Plasmon frequency $\omega_p$ is calculated at $q=1/d$. The first plot is for the Galilean invariant system and the second is for the generic case. 
  The sharp ridges on the plot correspond to dispersions of the symmetric and anisymmetric plasmon modes $\omega_\pm$. In the long wave length limit $x\ll1$ they disperse as $\propto\sqrt{x}$ and $\propto x$, merge  at $x\sim 1$, and gradually disappear at $x>1$. The spectral weight becomes small at $q\to0$ so resonances are not clearly visible. The dimensionless parameter $\beta$ controls the width of plasmon branches and is defined by Eqs. \eqref{eq:kappa-nu} and \eqref{eq:kappa-sigma-n} in each case respectively.} 
  \label{Fig-3D}
\end{figure}

\subsection{Dynamical structure factor}

It suffices to consider fluctuations in the stationary fluid $\bm{v}=0$. Note, however, that locally density fluctuations $\delta n$ trigger fluctuations of the velocity $\delta\bm{v}$ as it follows from the continuity. Thus we linearize the continuity equation \eqref{eq:dt-x} and find for the particle current fluctuations  
\begin{equation}\label{eq:j-linearized}
-i\omega\delta n+in(\bm{q}\cdot\delta\bm{v})+\frac{\sigma}{e^2}(q^2e\delta\Phi)+i(\bm{q}\cdot\bm{I}_{n})=0.
\end{equation} 
This equation applies to both layers. From the Navier-Stokes equation \eqref{eq:dt-p} we find
\begin{equation}\label{eq:p-linearized}
-i\rho\omega(\bm{q}\cdot\delta\bm{v})=-ienq^2\delta\Phi+i\bm{q}\cdot(\bm{q}\cdot\delta\Sigma),
\end{equation}
where we additionally multiplied by momentum $\bm{q}$ to have a scalar form of the equation. Here we neglected terms due to pressure fluctuations 
\begin{equation}\label{eq:delta-P}
\delta P=\left(\frac{\partial P}{\partial n}\right)_S\delta n+\left(\frac{\partial P}{\partial s}\right)_V\delta s.
\end{equation} 
Indeed, neglecting the first term is legitimate since density variations from the local compressibility are significantly smaller than the contribution to density fluctuations stemming from the long-range Coulomb potential, which is retained in Eq.  \eqref{eq:p-linearized}. The second term in Eq. \eqref{eq:delta-P} couples particle and entropy density fluctuations, which physically come from the diffusive modes caused by temperature fluctuations and thermal expansion of the electron liquid. Our analysis shows that these contributions are small as compared to plasmons driven by fluctuating longitudinal stresses in the fluid. The linearization of the stress tensor in Eq. \eqref{eq:stress} gives   
\begin{equation}
\bm{q}\cdot(\bm{q}\cdot\delta\Sigma)=i(\eta+\zeta)q^2(\bm{q}\cdot\delta\bm{v})+\bm{q}\cdot(\bm{q}\cdot\Xi). 
\end{equation}
Together with the form of the fluctuating potential in Eq. \eqref{eq:delta-Phi} we can eliminate velocity fluctuations $\delta\bm{v}$ from Eqs. \eqref{eq:j-linearized} and \eqref{eq:p-linearized} to arrive at two linear equations that describe coupled density fluctuations in each layer. In the symmetrized basis of normal modes these equations decouple and we find as a result 
\begin{subequations}
\begin{equation}
\delta n_\pm(\bm{q},\omega)=\delta n^\sigma_\pm(\bm{q},\omega)+\delta n^\nu_\pm(\bm{q},\omega),  
\end{equation}
\begin{equation}
\delta n^\sigma_\pm(\bm{q},\omega)=(\omega+i\omega_\nu)\frac{(\bm{q}\cdot\bm{I}_{n\pm})}{\Gamma_\pm(q,\omega)},
\end{equation}
\begin{equation}
\delta n^\nu_\pm(\bm{q},\omega)=(n/\rho)\frac{\bm{q}\cdot(\bm{q}\cdot\Xi_\pm)}{\Gamma_\pm(q,\omega)}.
\end{equation}
\end{subequations}
The splitting of density fluctuations into two separate contributions is natural since viscous stresses do not correlate with the intrinsic fluctuations of currents. The resonant denominator of the polarization function of electron fluid in a double layer is found in the form
\begin{equation}
\Gamma_\pm(q,\omega)=\omega^2-\omega^2_\pm-\chi_\pm\omega_\nu+i\omega(\omega_\nu+\chi_\pm),
\end{equation}
where 
\begin{equation}\label{eq:plasmons}
\omega^2_\pm=\omega^2_p(1\pm e^{-qd}),\quad \chi_\pm=\chi(1\pm e^{-qd}). 
\end{equation}
We introduced here several characteristic energy scales in the problem: viscous diffusion $\omega_\nu=\nu q^2$ with the kinematic viscosity $\nu=(\eta+\zeta)/\rho$, plasma frequency $\omega_p=\sqrt{2\pi(ne)^2q/\rho\epsilon}$, and Maxwellian relaxation rate $\chi=2\pi\sigma q/\epsilon$. Zeros of $\Gamma_\pm$ define dispersion relations of collective modes propagating in the system \cite{Forster}. If we set all dissipative kinetic coefficients to zero, then $\omega=\omega_\pm(q)$ gives us two low-lying gapless collective excitations which are plasmon poles as expected: $\omega_+\propto \sqrt{q}$ and $\omega_-\propto q$ as $q\to0$. The imaginary part of the plasmon dispersion defines attenuation of electron density oscillations. For example, in the Galilean invariant case $\Im\omega=\omega_\nu\propto q^2$; therefore, fluctuations with sufficiently low $q$ have long mean free path and as such plasmons remain well-defined excitation in the hydrodynamic regime. In the generic case of systems with broken Galilean invariance one finds instead that attenuation of plasmons is dominated by the Maxwell mechanism of charge relaxation, $\Im\omega=\chi_\pm$, due to finite intrinsic conductivity in the systems. This behavior should also be contrasted to the high-frequency (kinetic) regime, $\omega\gg v_Fq$, where plasmon attenuation is dominated by the decay into two particle-hole pairs. The corresponding rates for both Galilean invariant and Dirac systems scale as $\Im\omega\propto q^2$ \cite{Glazman,Maslov}.

To complete the derivation of the dynamic structure factor we need to perform thermal averages. This is easily done with the help of correlation functions defined by Eqs. \eqref{eq:zeta-zeta} and \eqref{eq:J-J}. After a simple algebra we find:
\begin{subequations}
\begin{equation}
\left\langle\bm{q}\cdot(\bm{q}\cdot\Xi_\pm)\bm{q}\cdot(\bm{q}\cdot\Xi_\mp)\right\rangle=4q^4(\eta+\zeta)  \Delta T,
\end{equation} 
\begin{equation}
\left\langle(\bm{q}\cdot\bm{I}_{n\pm})(\bm{q}\cdot\bm{I}_{n\mp})\right\rangle=2 q^2\frac{\sigma}{e^2} \Delta T.
\end{equation}
\end{subequations}
Using these relations we can express the density structure factor in Eq. \eqref{eq:D} in the following form
\begin{subequations}\label{eq:D-D}
\begin{equation}
D(\bm{q},\omega)=D^\nu(\bm{q},\omega)+D^\sigma(\bm{q},\omega),
\end{equation}
\begin{equation}
D^\nu(\bm{q},\omega)= 4\frac{\Gamma_+\Gamma^*_-}{|\Gamma_+|^2|\Gamma_-|^2}\frac{\omega_\nu(nq)^2}{\rho} \, \Delta T,
\end{equation}
\begin{equation}
D^\sigma(\bm{q},\omega)= 2 \frac{\Gamma_+\Gamma^*_-}{|\Gamma_+|^2|\Gamma_-|^2} \, (\omega^2+\omega^2_\nu) q^2\frac{\sigma}{e^2} \, \Delta T. 
\end{equation}
\end{subequations}
Here $D^\nu(\bm{q},\omega)$ denotes the contribution to the structure factor that arises from random thermal stresses, and $D^\sigma(\bm{q},\omega)$ denotes the contribution caused by random currents associated with the intrinsic conductivity. 
In the above expressions we made use of the obvious symmetry properties of $\Gamma_\pm(q,\omega)$, namely $\Gamma_\pm(q,\omega)=\Gamma_\pm (-q,\omega)$ and  $\Gamma_\pm(q,-\omega)=\Gamma^*_\pm(q,\omega)$. We also observe that the product $\Gamma_+\Gamma^*_-$ obeys the standard properties of response functions; its  imaginary part is odd in frequency and positive for positive frequencies. This makes the integrand in Eq.~\eqref{eq:W} sign-definite, resulting in the thermal conductance that is manifestly positively definite.

\begin{figure}[t!]
  \centering
  \includegraphics[width=1.65in]{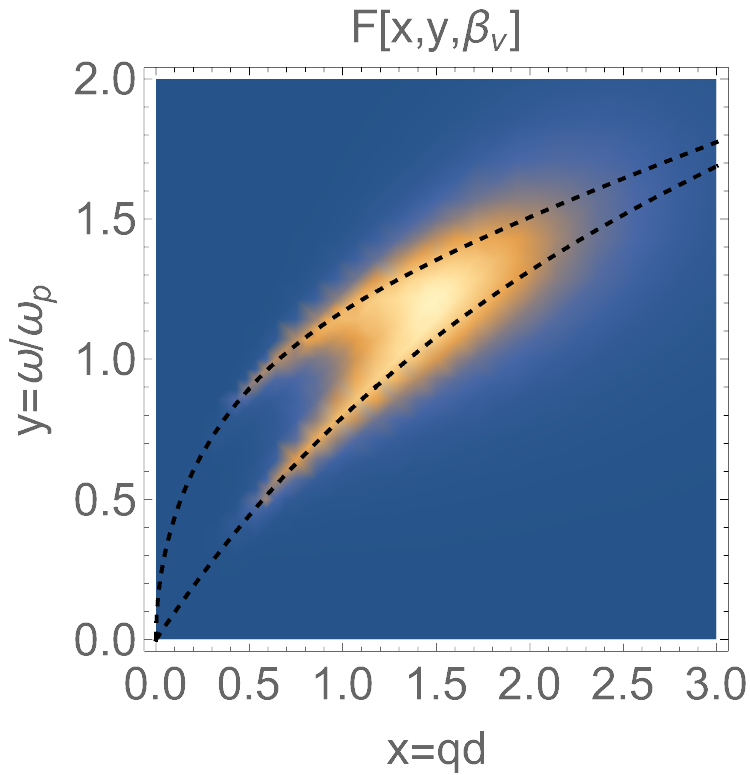}
  \includegraphics[width=1.65in]{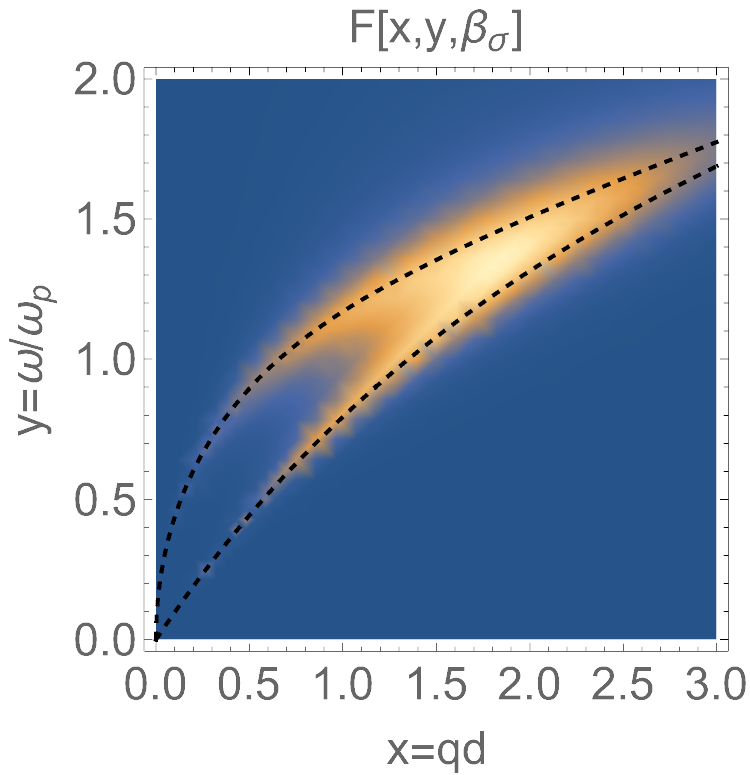}
  \caption{Plasmon dispersion laws $\omega=\omega_\pm(q)$ shown by dashed lines are superimposed on top of the color plot that defines contours of the $F(x,y,\beta)$ function from Fig. \ref{Fig-3D}. This plot is made for exactly the same choice of parameters as in Fig. \ref{Fig-3D}.}
  \label{Fig-2D}
\end{figure}

In order to highlight the importance of the plasmon resonance contribution to the heat flux we plotted the integrand of Eq. \eqref{eq:W} in the two-dimensional parameter spaces of frequencies and momenta. As is clear from the structure of Eq. \eqref{eq:W} this integrand is the product of the imaginary part of the dynamical structure factor, phase space volume, and Coulomb potential. The upper graph on Fig. \ref{Fig-3D} corresponds to the Galilean invariant case, whereas the lower plot is for the generic case. To generate these plots we normalized all momenta in units of interlayer separation and all frequencies in units of plasma frequency $\omega_p$ at $q=1/d$. In these dimensionless variables the integrand in each case depends on a single parameter  (defined later in the text) that controls the shape of the plot. In both cases plasmon branches are clearly visible and sharpest at $q\sim 1/d$. It is apparent from the plot that density fluctuations are strongly suppressed everywhere away from the plasmon resonances. This is further emphasized in the contrast plot of Fig. \ref{Fig-2D} that shows the same data as in Fig. \eqref{Fig-3D} projected to the $\omega-q$ plane.  

\begin{figure}[t!]
  \centering
  \includegraphics[width=3in]{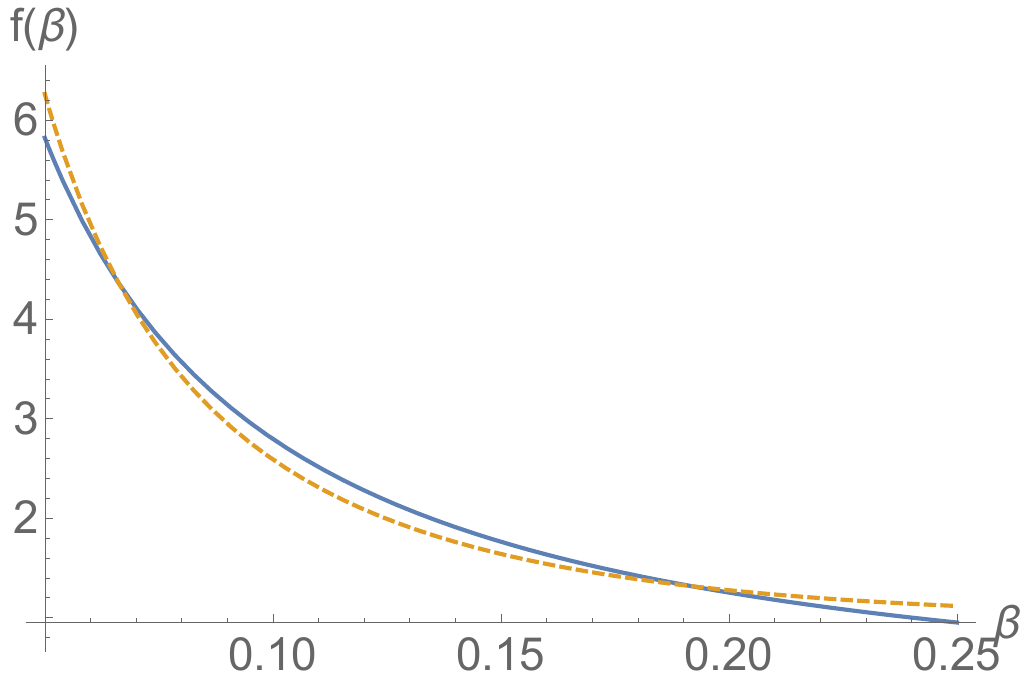}
  \caption{Plot of the dimensionless function $f(\beta)$ defined by Eq. \eqref{eq:f}. The solid line represents the result of numerical integration and the dashed line corresponds to the  approximate analytic formula in Eq. \eqref{eq:f-approx} applicable for $\beta<1$. }
  \label{Fig-f}
\end{figure}

\subsection{Galilean invariant systems}

We proceed to analyze the expression for the heat flux Eq. \eqref{eq:W} beginning with the case of Galilean invariant electron systems. We thus set $\sigma\to0$. Only the $D^\nu$ part of the structure factor from Eq. \eqref{eq:D-D} contributes in this case. From the definition of the NFET thermal conductance given in Eq. \eqref{eq:NFHT} we obtain 
\begin{equation}
\varkappa=\int\frac{d\omega d^2q}{(2\pi)^3}\left(\frac{4\pi e^2}{\epsilon \rho q}\right)e^{-qd}\frac{\omega^2_\nu\omega^2(\omega^2_+-\omega^2_-)(nq)^2}{|\Gamma_+|^2|\Gamma_-|^2}.
\end{equation}
The frequency integral can be evaluated analytically using the residue method,  
\begin{equation}
\int^{+\infty}_{-\infty}\frac{\omega^2d\omega}{|\Gamma_+|^2|\Gamma_-|^2}=\frac{2\pi/\omega_\nu}{(\omega^2_+-\omega^2_-)^2+2\omega^2_\nu(\omega^2_++\omega^2_-)}.
\end{equation}
We then use the explicit forms of the dispersion relations Eq. \eqref{eq:plasmons} to simplify the remaining momentum integral. Introducing dimensionless variable $x=qd$ we arrive at the result 
\begin{equation}\label{eq:kappa-nu}
\varkappa=\frac{\nu}{2\pi d^4}f(\beta_\nu),\quad \beta_\nu=\frac{\nu}{d^2\omega_p}.
\end{equation} 
It should be understood that in the definition of the parameter $\beta_\nu$ the plasma frequency $\omega_p$ is taken at $q=1/d$. The dimensionless  function $f(\beta)$ is defined by the following integral 
\begin{equation}\label{eq:f}
f(\beta)=\int^{\infty}_{0}\frac{x^3e^{-2x}dx}{e^{-2x}+\beta^2x^3}.
\end{equation} 
In the physically relevant regime of parameters $\beta_\nu\ll1$. We are able to extract an asymptotic form of $f(\beta)$ in that limit. With the logarithmic accuracy we find
\begin{equation}\label{eq:f-approx}
f(\beta)\approx\Lambda\ln\left(\frac{1}{\beta\sqrt{\Lambda}}\right),\quad \Lambda=\ln^3\left(\frac{1}{\beta\ln^{3/2}(1/\beta)}\right). 
\end{equation}  
To verify the validity of this approximation we compare it with the result of numerical integration in Fig. \ref{Fig-f}.  

\begin{figure}[t!]
  \centering
  \includegraphics[width=3in]{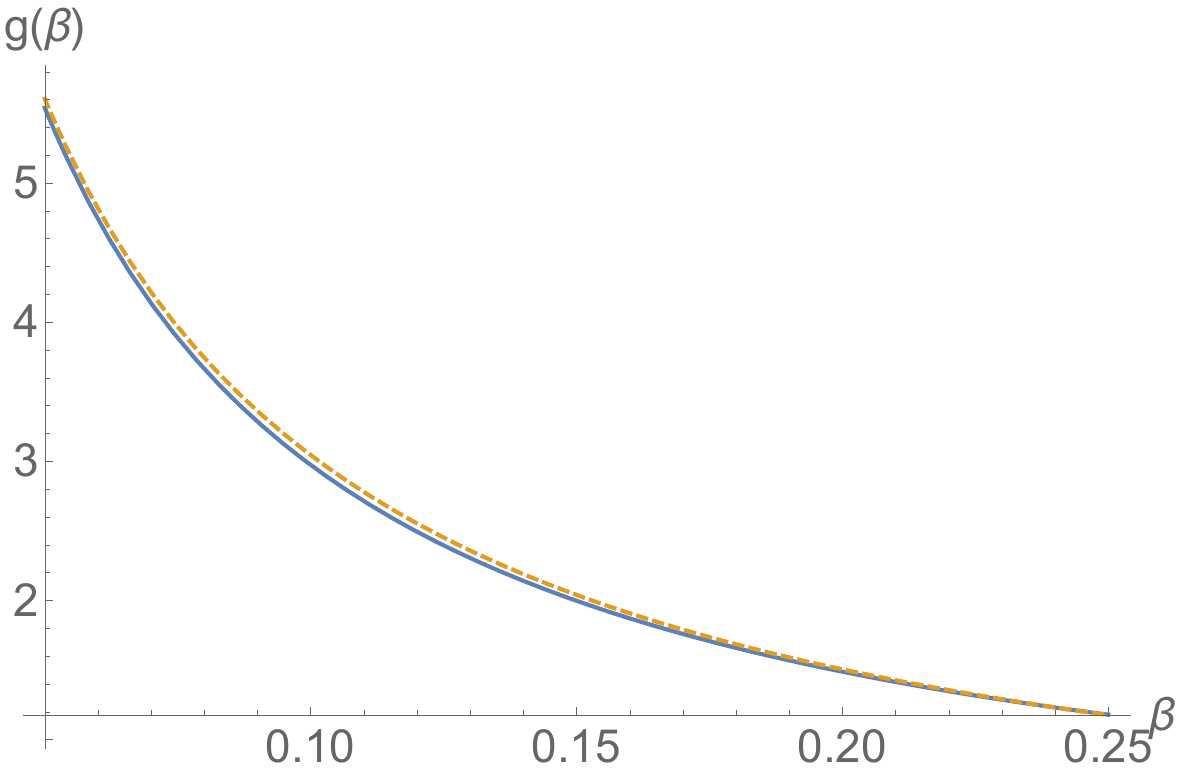}
  \caption{Plot of the dimensionless function $g(\beta)$ defined by Eq. \eqref{eq:kappa-sigma-n}. The solid line represents the result of numerical integration, whereas the dashed line corresponds to the approximate analytical expression in Eq. \eqref{eq:g-approx} applicable for $\beta<1$. }
  \label{Fig-g}
\end{figure}

\subsection{Systems with broken Galilean invariance}

In generic systems, where both viscosity and intrinsic conductivity of the electron liquid are nonzero, the density fluctuations responsible for NFET are generated not only by random stresses but also by random intrinsic currents. We begin by examining the latter contribution, as it is found to be dominant, and comment on the modification to the viscous contribution later on. The discussion presented in this section pertains to double-layer devices consisting of, e.g., graphene monolayer or bilayer sheets.  

Using $D^\sigma$ from Eq. \eqref{eq:D-D} in the expression for the energy flux Eq. \eqref{eq:W} we obtain the following expression for the contribution of intrinsic conductivity to the thermal  conductance: 
\begin{equation}\label{eq:kappa-sigma}
\varkappa=\frac{\sigma}{e^2}\int\frac{d\omega d^2q}{(2\pi)^3}\left(\frac{2\pi e^2}{\epsilon q}\right)e^{-qd}\frac{q^2(\omega^2+\omega^2_\nu)\omega\Im(\Gamma_+\Gamma^*_-)}{|\Gamma_+|^2|\Gamma_-|^2}
\end{equation}

Let us first focus on the limit of charge neutrality, $n\to0$. In this case the expressions simplify greatly; $\Gamma_\pm\approx \omega(\omega+i\chi_\pm)$ and $\Im(\Gamma_+\Gamma^*_-)\approx \omega^3(\chi_+-\chi_-)$. Notice that the dynamics of charge fluctuations is overdamped and corresponds to Maxwell charge relaxation with the intrinsic conductivity of the electron liquid. The important point that there are no hydrodynamic plasmons in undoped graphene was thoroughly discussed in the literature recently \cite{Titov,Stoof}. Using an integral 
\begin{equation}
\int^{+\infty}_{-\infty}\frac{\omega^2d\omega}{(\omega^2+\chi^2_+)(\omega^2+\chi^2_-)}=\frac{\pi}{\chi_++\chi_-},
\end{equation}    
we arrive at the following expression 
\begin{equation}
\varkappa=\frac{\sigma}{e^2}\int\frac{q^2d^2q}{(2\pi)^2}\left(\frac{\pi e^2}{\epsilon q}\right)e^{-qd}\frac{\chi_+-\chi_-}{\chi_++\chi_-}.
\end{equation}
Note that the $\omega^2_\nu$ term in the numerator of Eq.  \eqref{eq:kappa-sigma} can be neglected as it contains four more powers of $q$, while each power brings a factor of $1/d$ after the momentum integration. Therefore, this piece will be subleading in powers of $1/d$ as compared to the main contribution. As the final step, using explicit forms of $\chi_\pm$ from Eq.~\eqref{eq:plasmons}, we find after the remaining momentum integration 
\begin{equation}\label{eq:kappa-sigma-CNP}
\varkappa=\frac{\sigma}{8\epsilon d^3}. 
\end{equation}
This is in agreement with the earlier analysis of NFET at the dual neutrality point \cite{Levchenko}. 

Given the degree of tunability of graphene devices we investigate further how NFET depends on the carrier concentration. In order to analyze the high doping regime we can approximate $\Gamma_\pm\approx \omega^2-\omega^2_\pm+i\omega\chi_\pm$ and notice that $\Im(\Gamma_+\Gamma^*_-)\approx\omega^3(\chi_+-\chi_-)$ still holds since $\chi_-\omega^2_+-\chi_+\omega^2_-=0$. While the frequency integral in  Eq. \eqref{eq:kappa-sigma} can be evaluated analytically the resulting expression is too cumbersome to present here.  We establish the final result in the form 
\begin{equation}\label{eq:kappa-sigma-n}
\varkappa=\frac{\sigma}{\pi\epsilon d^3}g(\beta_\sigma),\quad \beta_\sigma=\frac{\chi}{\omega_p},
\end{equation}
where the dimensionless function $g(\beta)$ can be efficiently computed numerically; see Fig. \ref{Fig-g}. In the limit $\beta_\sigma\ll1$ a rather accurate estimate can be found analytically:
\begin{equation}\label{eq:g-approx}
g(\beta)\approx\frac{\pi\Lambda}{16}\ln\left(\frac{\sqrt{2}}{\beta\sqrt[4]{\Lambda}}\right),\quad \Lambda=\ln^2\left(\frac{2}{\beta\ln^{1/2}(1/\beta)}\right).
\end{equation} 
We see that as compared to Eq. \eqref{eq:kappa-sigma-CNP} the intrinsic contribution has only a modest (logarithmic) density dependence encoded by a parameter $\beta_\sigma$.  

The viscous contribution is also present in this case. The analysis based on the form of $D^\nu$ shows that $\varkappa$ is given by the expression similar to Eq. \eqref{eq:kappa-nu} with the difference, $f(\beta_\nu)\to h(\beta_\sigma)$, where $h(\beta)$ is yet another logarithmically slow dimensionless function. In  view of the fact that this viscous term scales $\propto 1/d^4$ it remains parametrically smaller than the contribution of intrinsic conductivity given by Eq. \eqref{eq:kappa-sigma-n}.   

\subsection{Discussion}

Let us now obtain estimates for NFET thermal conductance $\varkappa(T)$ in the hydrodynamic regime and discuss their implications for the general picture of temperature dependence on NFET thermal conductance in high mobility electron double layers. 

\subsubsection{Temperature regimes}

As alluded above, the hydrodynamic description applies at intermediate-to-high temperatures where the intralayer mean free path limited by electron-electron scattering is short as compared to  interlayer distance $d$. For Fermi liquids with the typical rate of electron-electron scattering given by $\gamma_{ee}\propto T^2/E_F$, where $E_F$ is the Fermi energy, we have $T>T_c\sim E_F/\sqrt{k_Fd}$ with $k_F$ being Fermi momentum. The scale of $T_c$ marks the onset of the collision-dominated regime for the relevant density fluctuations in the particle-hole continuum.  The crossover to the collision-dominated regime for plasmons, $\gamma_{ee}>\omega_p$, occurs at a higher temperature scale, $T_h\sim\sqrt{E_F\omega_p}\sim E_F/\sqrt[4]{k_Fd}$. The consideration of the crossover regime, $T_c<T<T_h$, is beyond the scope of this work. It requires a more detailed theory based on the kinetic equation, so-called Boltzmann-Langevin approach \cite{Kogan}. The contribution of the particle-hole continuum to the NFET thermal conductance at lower temperatures, $T<T_c$, is comprehensively covered in the literature; see, for example, Ref. \cite{Kamenev}. Thus we focus on temperatures $T_h<T< E_F$. 

\subsubsection{Galilean invariant systems}

Recall that shear viscosity of the Fermi liquid is $\eta\sim n (E_F/T)^2$ \cite{Lyakhov}, where logarithmic corrections specific to the 2D case were disregarded for brevity. Therefore, based on Eq. \eqref{eq:kappa-nu} we conclude that NFET thermal conductance diminishes as $\propto 1/T^2$. In contrast, at the lowest temperatures where the thermal charge fluctuations responsible for NFET correspond to electron-hole pairs in the collisionless regime, the interlayer thermal conductance increases with temperature. Thus the overall temperature dependence of $\varkappa$ is nonmonotonic.  We expect $\varkappa$ to have a peak at $T\sim T_h$ as dominated by plasmons, which can be easily estimated from Eq. \eqref{eq:kappa-nu}  
\begin{equation}\label{eq:kappa-estimate-GI}
\varkappa(T_h) \sim E_Fk^2_F(k_Fd)^{-7/2}. 
\end{equation}
In the same range of parameters we also estimate that parameter $\beta_\nu$ that enters Eq. \eqref{eq:kappa-nu} is of the order 
\begin{equation}
\beta_\nu\sim\sqrt{\frac{\epsilon v_F}{e^2}}\frac{1}{k_Fd}\ll1. 
\end{equation}
This estimate justifies approximations made in the derivation of the asymptote of $f(\beta)$ function in Eq. \eqref{eq:f-approx}. 

\subsubsection{Graphene devices}

In generic systems with broken Galilean invariance temperature dependence of the NFET thermal conductance is primarily governed by the intrinsic conductivity $\sigma(T)$ at all densities. For instance, for graphene monolayers --- Dirac fluid near charge neutrality --- the known results \cite{Mishchenko,Kashuba,Fritz} of the perturbative renormalization group give an estimate 
\begin{equation}
\sigma(T)\sim \sigma_Q\ln^2\frac{U}{T}, \quad n\to0,
\end{equation} 
where $\sigma_Q$ is the conductance quantum and $U$ is a cutoff energy scale of the order of the electronic bandwidth. Somewhat surprisingly, the presence of the plasmon resonances at a nonzero density leads only to  modest logarithmic modifications of the result as compared to an undoped limit [per Eq. \eqref{eq:kappa-sigma-n}]. The parameter $\beta_\sigma$ in Eq. \eqref{eq:kappa-sigma-n} that determines the enhancement factor in Eq. \eqref{eq:g-approx} can be estimated to be of the order of
\begin{equation}
\beta_\sigma\sim\frac{\sigma}{e^2}\sqrt{\frac{e^2}{\epsilon v_F}}\frac{1}{\sqrt{k_Fd}}\ll1.
\end{equation}
Consequently, for a double-layer system comprised of graphene monolayers at nonzero density, the comparison of the estimate of Eq. \eqref{eq:kappa-sigma-n},
\begin{equation}
\varkappa(T_h)\sim E_Fk^2_F(\sigma/\epsilon v_F)(k_Fd)^{-3}, 
\end{equation}
with  Eq. \eqref{eq:kappa-estimate-GI}, shows that the plasmon enhancement of NFET is enhanced by the parameter  $\sqrt{k_Fd}\gg1$, as compared to the Galilean invariant systems. 

\section{Summary}\label{sec:summ}

In this work we developed a theory of near-field energy transfer between two-dimensional electron systems in the hydrodynamic regime. 
In this regime the interlayer thermal conductance is expressed in terms of the dissipative coefficients of the electron liquid, namely it's viscosity and (in the absence of Galilean invariance) intrinsic conductivity. Applicability of the hydrodynamic description requires that the range of temperatures, particle density, and carrier mobility are such that the electron mean free path is the shortest length scale in the problem. These conditions can be realized in modern semiconductor quantum-well heterostructures and van der Waals materials.  We considered both Galilean invariant liquids and electron systems where this invariance is broken, such as in graphene devices. 

Using Ehrenfest's theorem, we obtained a general formula for the NFET conductance in terms of the dynamical structure factor of the fluid, Eq. \eqref{eq:W}. We then obtained the hydrodynamic expression for NFET thermal conductance by evaluating the contribution of hydrodynamic thermal fluctuations to the structure factor. We found that plasmon resonances produce a strong enhancement of NFET. We note that temperature fluctuations can also be described in this framework, but produce a subleading effect because they correspond to diffusive spreading at quasineutrality. In Galilean-invariant systems NFET conductance $\varkappa$, as well as the plasmon lifetime, is determined by the kinematic viscosity $\nu$ of the electron fluid. Modulo logarithmic corrections [see Eq. \eqref{eq:kappa-nu}] $\varkappa$ falls off with interlayed distance as $1/d^4$
and follows the same temperature dependence as $\nu$. In systems without Galilean invariance charge fluctuations relax exponentially with the Maxwell rate [see Eq. \eqref{eq:kappa-sigma}]. As a result, $\varkappa$ is governed by the intrinsic conductivity of the electron liquid and falls off as $1/d^3$ with interlayer distance.   
The strong NFET effect survives even at charge neutrality where plasmons are absent.        
At finite density this effect persists with further enhancement by plasmons [Eq. \eqref{eq:kappa-sigma-n}]. 

The nonperturbative character of our results makes them applicable to strongly correlated systems. 
We believe that NFET effect could prove fruitful in further exploration of strongly interacting electron systems, including anomalous metals \cite{Kapitulnik}, strongly correlated 2D systems \cite{Spivak}, and  strange metals \cite{Chudnovskiy}.    

\section*{Acknowledgments}

We thank L. Glazman and D. Maslov for the discussion of plasmon attenuation in Fermi and Dirac electron liquids per Refs. \cite{Glazman,Maslov}. This work at the UW-Madison was supported by the U.S. Department of Energy (DOE), Office of Science, Basic Energy Sciences (BES) under Award No. DE-SC0020313.  The work of A.V.A. was supported by the US National Science Foundation through the MRSEC Grant No. DMR-1719797, the Thouless Institute for Quantum Matter (TIQM), and the College of Arts and Sciences at the University of Washington. A. L. acknowledges the hospitality of TIQM during the Winter Workshop 2023 and the support during the stay.

\bibliography{biblio}

\end{document}